\documentclass[conference]{IEEEtran}
\usepackage{cite}
\usepackage{amsmath,amssymb,amsfonts}
\usepackage{algorithmic}
\usepackage{graphicx}
\usepackage{textcomp}
\usepackage{xcolor}
\usepackage{spconf}
\usepackage{bm}
\usepackage{algorithm}
\usepackage[caption=false,font=footnotesize]{subfig}
\usepackage{float}

\def\BibTeX{{\rm B\kern-.05em{\sc i\kern-.025em b}\kern-.08em
    T\kern-.1667em\lower.7ex\hbox{E}\kern-.125emX}}

\DeclareMathOperator{\tr}{\text{tr}}

\DeclareMathOperator{\cov}{\text{Cov}}
\DeclareMathOperator{\var}{\text{Var}}

\DeclareGraphicsExtensions{.eps}

\begin{document}

\title{Active Sampling for Approximately Bandlimited Graph Signals}

\name{Sijie Lin, Xuan Xie, Hui Feng, Bo Hu}
\address{Research Center of Smart Networks and Systems, School of Information Science and Engineering\\
Fudan University, Shanghai 200433, China\\
Emails: \{sjlin18, xxie15, hfeng, bohu\}@fudan.edu.cn}

\maketitle

\begin{abstract}

This paper investigates the active sampling for estimation of approximately bandlimited graph signals. With the assistance of a graph filter, an approximately bandlimited graph signal can be formulated by a Gaussian random field over the graph. In contrast to offline sampling set design methods which usually rely on accurate prior knowledge about the model, unknown parameters in signal and noise distribution are allowed in the proposed active sampling algorithm. The active sampling process is divided into two alternating stages: unknown parameters are first estimated by Expectation Maximization (EM), with which the next node to sample is selected based on historical observations according to predictive uncertainty. Validated by simulations compared with related approaches, the proposed algorithm can reduce the sample size to reach a certain estimation accuracy.
\end{abstract}


\section{Introduction}
\label{sec1 introduction}

Sampling is a fundamental problem in graph signal processing \cite{shuman2013emerging,ortega2018graph}. For some large networks, it is costly or impractical to acquire the exact signal value on each node. Instead, the entire signal has to be recovered or estimated from the observations on a portion of nodes under smoothness assumption. There has been plenty of research on the sampling of bandlimited \cite{pesenson2008sampling,narang2013signal,anis2014towards,chen2015discrete,marques2016sampling,xie2017design,anis2016efficient} or approximately bandlimited \cite{anis2016efficient,chen2016signal,perraudin2018global} graph signals, including noise-free sampling for reconstruction \cite{pesenson2008sampling,narang2013signal,anis2014towards,chen2015discrete,anis2016efficient} and noisy sampling for estimation \cite{marques2016sampling,xie2017design,perraudin2018global,anis2016efficient,chen2016signal}.

Focusing on the scenario with sampling noise, from a statistical view, optimal sample selection is to minimize the estimation error in expectation. In another perspective, graph signal sampling can also be regarded as pool-based active semi-supervised learning. Current sampling methods are mostly offline \cite{marques2016sampling,xie2017design}, in which sampling set is designed in advance taking into account only graph structure. These approaches are usually not applicable without accurate prior knowledge about the distribution of signal and noise. To reduce the reliance on prior knowledge, we try to introduce online active learning \cite{settles2012active} to the sampling of graph signals, turning sampling into a sequential decision process. Involving historical observations, unknown parameters of the underlying model can be gradually estimated in the process of sampling, and subsequent samples can be chosen based on the latest estimation results.

In fact, online active learning has been applied to graph-aware classification \cite{zhu2003combining,long2008graph,zhou2014active,berberidis2017active}. These active sampling algorithms select samples online taking into account both graph structure and previously obtained labels, yet they can not apply directly to continuous graph signals. \cite{anis2016efficient} considers the active semi-supervised learning for continuous graph signals, yet it is offline batch-mode without using any label information.

In this paper, the online active sampling of continuous graph signals with sampling noise is considered under a Bayesian framework. We propose an active sampling algorithm for approximately bandlimited graph signals where the exact statistical properties of the signal and noise are unknown. Parameter estimation and node selection are integrated into a unified Bayesian framework. At each step, EM algorithm \cite[Sec.11.2]{barber2012bayesian} is first used to estimate the unknown parameters in the signal and noise distribution. Then the node with largest predictive variance, that is, the node whose value we are most uncertain about, is selected to be sampled at the next step. Making full use of historical observations, the proposed algorithm is expected to reduce the number of samples required to reach a certain estimation accuracy.


\section{System Model}
\label{sec2 system model}
Consider a time-invariant signal $\bm{f}\in\mathbb{R}^N$ defined on an $N$-vertex undirected connected graph $\mathcal{G}$. From a Bayesian view, $\bm{f}$ is \emph{approximately bandlimited} formulated by the following Gaussian random field over the graph
\begin{equation}
\label{prior}
p\left(\bm{f}\,|\,\alpha\right)=\mathcal{N}\left(\bm{0},\alpha^{-1}\bm{H}^{-2}\right)\propto\exp\left(-\dfrac{\alpha}{2}\bm{f^T\bm{H}^2\bm{f}}\right),
\end{equation}
where $\bm{H}$ is a unit-gain high-pass graph filter \cite{shuman2013emerging}, and $\alpha$ is a parameter.

This is a generalized form of that in \cite{gadde2015probabilistic}. $\bm{H}$ can be an FIR \cite{shuman2011chebyshev} or IIR \cite{shi2015infinite} filter, the cut-off frequency of which can be regarded as the approximate bandwidth of $\bm{f}$. The term $\bm{f}^T\bm{H}^2\bm{f}=\Vert\bm{Hf}\Vert_2^2$ represents the energy of out-of-band components of the signal. According to (\ref{prior}), larger $\bm{f}^T\bm{H}^2\bm{f}$ results in lower probability, restricting the high-frequency energy of the signal to a relatively low level.

As for $\alpha$, the larger $\alpha$ is, the more rapidly the probability drops as $\bm{f}^T\bm{H}^2\bm{f}$ increases. In some sense, $\alpha$ governs the approximation degree of $\bm{f}$ to a bandlimited graph signal, and is relevant to the smoothness in the vertex domain.

In contrast to batched sampling, here we consider an active sampling process where only one node is sampled at each time. Suppose that at time $t$, the $n$-th node is sampled and the observed signal value is $y(t)$. The observation model at time $t$ is
\begin{equation}
\label{observation model 1}
y(t)=\bm{\psi}^T(t)\bm{f}+w(t)\:,
\end{equation}
where $\bm{\psi}^T(t)$ is the sampling vector, a row vector with the $n$-th entry equal to 1 and the others equal to 0, and $w(t)$ is Gaussian sampling noise with zero mean and precision $\beta$
\begin{equation}
\label{noise 1}
p\left(w(t)\,|\,\beta\right)=\mathcal{N}\left(0,\beta^{-1}\right).
\end{equation}

Let vector $\bm{y}_s(t)=\left(y(1),\:\cdots,\:y(t)\right)^T$ be the historical observations up to time $t$. According to (\ref{observation model 1}),
\begin{equation}
\label{observation model M}
\bm{y}_s(t)=\bm{\Psi}(t)\bm{f}+\bm{w}(t)\:,
\end{equation}
where $\bm{\Psi}(t)$ is a $M\times N$ sampling matrix whose rows are sampling vectors $\bm{\psi}^T(1),\:\cdots,\:\bm{\psi}^T(t)$, and sampling noise $\bm{w}(t)$ follows a joint distribution
\begin{equation}
\label{noise M}
p\left(\bm{w}(t)\,|\,\beta\right)=\mathcal{N}\left(\bm{0},\beta^{-1}\bm{I}\right).
\end{equation}

In this paper, high-pass graph filter $\bm{H}$ in the signal prior (\ref{prior}) is assumed to be given, while parameter $\alpha$ and noise precision $\beta$ are considered unknown.

Under this framework, the core problem is: at each time $t$, how to estimate $\alpha$ and $\beta$ and then decide the next node to sample (at time $t+1$) based on historical observations $\bm{y}_s(t)$, in order to better estimate the signal $\bm{f}$ with less samples.

\section{Algorithm}
\label{sec3 algorithm}
At each time $t$, the next node to sample is decided based on historical observations $\bm{y}_s(t)$ (and $\bm{\Psi}(t)$). According to the \emph{uncertainty sampling} criterion in active learning\cite[Ch.2]{settles2012active}, a reasonable strategy is to evaluate the predictive distribution $p\left(y\,|\,\bm{\psi}^T,\bm{\Psi}(t),\bm{y}_s(t)\right)$ with different sampling vectors $\bm{\psi}^T$, and then select the node with maximum predictive variance to sample at next step.

Unknown parameters $\alpha$ and $\beta$, both of great importance for estimation and prediction, are updated sequentially using the EM algorithm \cite[Sec.11.2]{barber2012bayesian}. 

\subsection{Signal Estimation and Prediction}
\label{subsecA}
We first discuss the estimation of $\bm{f}$ and the prediction of the observed signal value on each node based on historical observations under a Bayesian framework, which form the basis for hyperparameter estimation and sample selection. For brevity, the time index $(t)$ of $\bm{\Psi}$, $\bm{y}_s$ are omitted.

Given $\alpha$, $\beta$, $\bm{\Psi}$ and $\bm{y}_s$, the posterior distribution of $\bm{f}$ can be calculated by
\begin{equation}
p\left(\bm{f}\,|\,\bm{\Psi},\bm{y}_s,\alpha,\beta\right)=\dfrac{p\left(\bm{y}_s\,|\,\bm{\Psi},\bm{f},\beta\right)p\left(\bm{f}\,|\,\alpha\right)}{\int p\left(\bm{y}_s\,|\,\bm{\Psi},\bm{f},\beta\right)p\left(\bm{f}\,|\,\alpha\right)d\bm{f}}\:.
\end{equation}
The signal prior $p\left(\bm{f}\,|\,\alpha\right)$ is Gaussian as given in (\ref{prior}), and making use of (\ref{observation model M}) and (\ref{noise M}), the likelihood $p\left(\bm{y}_s\,|\,\bm{\Psi},\bm{f},\beta\right)$ is also Gaussian
\begin{equation}
\label{likelihood M}
p\left(\bm{y}_s\,|\,\bm{\Psi},\bm{f},\beta\right)=\mathcal{N}\left(\bm{\Psi}\bm{f},\beta^{-1}\bm{I}\right).
\end{equation}
Consequently, the posterior distribution of $\bm{f}$ takes a Gaussian form, with mean and covariance\cite[Sec.10.6]{kay1993fundamentals}
\begin{align}
\label{posterior mean}
&\,\mathbb{E}\left[\bm{f}\,|\,\bm{\Psi},\bm{y}_s,\alpha,\beta\right]=\beta\left(\alpha\bm{H}^2+\beta\bm{\Psi}^T\bm{\Psi}\right)^{-1}\bm{\Psi}^T\bm{y}_s\:,\\
\label{posterior covariance}
&\cov\left[\bm{f}\,|\,\bm{\Psi},\bm{y}_s,\alpha,\beta\right]=\left(\alpha\bm{H}^2+\beta\bm{\Psi}^T\bm{\Psi}\right)^{-1}.
\end{align}

Subsequently, we can predict the observed signal value on each node by evaluating the predictive distribution
\begin{align}
&p\left(y\,|\,\bm{\psi}^T,\bm{\Psi},\bm{y}_s,\alpha,\beta\right)\nonumber\\
=&\int\! p\left(y\,|\,\bm{\psi}^T,\bm{f},\beta\right)p\left(\bm{f}\,|\,\bm{\Psi},\bm{y}_s,\alpha,\beta\right)d\bm{f},
\end{align}
in which $p\left(\bm{f}\,|\,\bm{\Psi},\bm{y}_s,\alpha,\beta\right)$ is the posterior distribution (\ref{posterior mean})-(\ref{posterior covariance}), and according to (\ref{observation model 1}) and (\ref{noise 1}),
\begin{equation}
\label{likelihood 1}
p\left(y\,|\,\bm{\psi}^T,\bm{f},\beta\right)=\mathcal{N}\left(\bm{\psi}^T\bm{f},\beta^{-1}\right).
\end{equation}
Thus the predictive distribution is still Gaussian, with mean and variance
\begin{align}
\label{predictive mean}
&\mathbb{E}\left[y\,|\,\bm{\psi}^T,\bm{\Psi},\bm{y}_s,\alpha,\beta\right]=\bm{\psi}^T\,\mathbb{E}\left[\bm{f}\,|\,\bm{\Psi},\bm{y}_s,\alpha,\beta\right],\\
\label{predictive variance}
&\!\var\left[y\,|\,\bm{\psi}^T,\bm{\Psi},\bm{y}_s,\alpha,\beta\right]=\bm{\psi}^T\cov\left[\bm{f}\,|\,\bm{\Psi},\bm{y}_s,\alpha,\beta\right]\bm{\psi}\nonumber\\
&\ \ \ \ \ \ \ \ \ \ \ \ \ \ \ \ \ \ \ \ \ \ \ \ \ \ \ \ \ \ \ +\beta^{-1}.
\end{align}

\subsection{Uncertainty Sampling}
\label{subsecB}
Consider the situation where any node is allowed to be sampled multiple times. According to the \emph{uncertainty sampling} criterion \cite[Ch.2]{settles2012active}, we should scan through all the nodes, and sample the one whose observed signal value is in greatest uncertainty.

In the proposed method, predictive variance is regarded as a measurement of uncertainty. Thus, the sampling vector at time $t+1$ is designed by
\begin{align}
\label{uncertainty sampling}
\bm{\psi}^T(t+1)&=\underset{\bm{\psi}^T\in\mathcal{S}}{\text{argmax}}\var\left[y\,|\,\bm{\psi}^T,\bm{\Psi}(t),\bm{y}_s(t),\alpha,\beta\right],
\end{align}
where $\mathcal{S}=\left\lbrace\bm{e}_n^T,\:n=1,2,\cdots,N\right\rbrace$ is the set of all possible sampling vectors.

In implementation, (\ref{uncertainty sampling}) can be achieved by searching for the greatest diagonal element in the posterior covariance matrix (6), which reduces the computational complexity of our algorithm.

By employing this method, we can avoid sampling the nodes whose value we are already confident about, and focus our attention on only those we find confusing.

\subsection{Estimation Using EM}
As illustrated above, unknown parameters $\alpha$ and $\beta$ are of major significance in sample selection. So every time before choosing the next sampling node, $\alpha$ and $\beta$ should first be re-estimated based on historical data.

A naive idea is to estimate these hyperparameters by maximizing the likelihood $p\left(\bm{y}_s(t)\,|\,\bm{\Psi}(t),\alpha,\beta\right)$. However, direct optimization is intractable. EM algorithm \cite[Sec.11.2]{barber2012bayesian} is introduced in our method to estimate $\hat{\alpha}$ and $\hat{\beta}$, and thereby obtain the posterior distribution of the signal.

Still for brevity, time indices $(t)$ are omitted. And let $k$ be the iteration index inside EM.

In E step, we fix $\hat{\alpha}$ and $\hat{\beta}$, and update the posterior distribution of $\bm{f}$ as
\begin{equation}
\label{E step}
q_k\left(\bm{f}\,|\,\bm{\Psi},\bm{y}_s\right)=p\left(\bm{f}\,|\,\bm{\Psi},\bm{y}_s,\hat{\alpha}_{k-1},\hat{\beta}_{k-1}\right).
\end{equation}
According to (\ref{posterior mean})-(\ref{posterior covariance}),
\begin{align}
\label{E mean}
\bm{\mu}_{k-1}&\triangleq\mathbb{E}\left[\bm{f}\,|\,\bm{\Psi},\bm{y}_s,\hat{\alpha}_{k-1},\hat{\beta}_{k-1}\right]\nonumber\\
&=\hat{\beta}_{k-1}\left(\hat{\alpha}_{k-1}\bm{H}^2+\hat{\beta}_{k-1}\bm{\Psi}^T\bm{\Psi}\right)^{-1}\bm{\Psi}^T\bm{y}_s\:,\\
\label{E covariance}
\bm{C}_{k-1}&\triangleq\cov\left[\bm{f}\,|\,\bm{\Psi},\bm{y}_s,\hat{\alpha}_{k-1},\hat{\beta}_{k-1}\right]\nonumber\\
&=\left(\hat{\alpha}_{k-1}\bm{H}^2+\hat{\beta}_{k-1}\bm{\Psi}^T\bm{\Psi}\right)^{-1}.
\end{align}

In M step, $\hat{\alpha}$ and $\hat{\beta}$ are updated by
\begin{equation}
\label{M step}
\hat{\alpha}_k,\hat{\beta}_k=\underset{\alpha,\beta}{\text{argmax}}\:\mathbb{E}_{\bm{f}|\bm{\Psi},\bm{y}_s,\hat{\alpha}_{k-1},\hat{\beta}_{k-1}}\left[\ln p\left(\bm{f},\bm{y}_s\,|\,\bm{\Psi},\alpha,\beta\right)\right],
\end{equation}
where according to the product rule of probability,
\begin{equation}
p\left(\bm{f},\bm{y}_s\,|\,\bm{\Psi},\alpha,\beta\right)=p\left(\bm{y}_s\,|\,\bm{\Psi},\bm{f},\beta\right)p\left(\bm{f}\,|\,\alpha\right).
\end{equation}
By (\ref{prior}) and (\ref{likelihood M}), we have
\begin{align}
\ln p\left(\bm{f},\bm{y}_s\,|\,\bm{\Psi},\alpha,\beta\right)=\:&\dfrac{N}{2}\ln\alpha+\dfrac{M}{2}\ln\beta-\dfrac{\alpha}{2}\bm{f}^T\bm{H}^2\bm{f}\nonumber\\
&-\dfrac{\beta}{2}\left(\bm{y}_s-\bm{\Psi f}\right)^T\left(\bm{y}_s-\bm{\Psi f}\right)\nonumber\\
&+\text{const},
\end{align}
where $M=t$ is the current sample size, and 'const' denotes terms that are independent of $\alpha$ and $\beta$.

It is obvious that the target function in (\ref{M step}) is concave with respect to $\alpha$ and $\beta$. To maximize it, take the partial derivatives and set them to zero. The M-step updates of $\hat{\alpha}$ and $\hat{\beta}$ are
\begin{align}
\label{M alpha}
\hat{\alpha}_k&=\dfrac{N}{\left(\tr\left(\bm{H}^2\bm{C}_{k-1}\right)+\bm{\mu}_{k-1}^T\bm{H}^2\bm{\mu}_{k-1}\right)},\\
\label{M beta}
\hat{\beta}_k&=\dfrac{M}{\left(\left(\bm{y}_s-\bm{\Psi}\bm{\mu}_{k-1}\right)^T\!\left(\bm{y}_s-\bm{\Psi}\bm{\mu}_{k-1}\right)+\tr\left(\bm{\Psi}^T\bm{\Psi}\bm{C}_{k-1}\right)\right)}.
\end{align}

Repeat the E and M steps until convergence, and we will obtain the maximum likelihood (ML) estimation of $\alpha$ and $\beta$, and the latest posterior distribution of $\bm{f}$ as well.

Then we can implement the sampling strategy in subsection \ref{subsecA} and \ref{subsecB} by replacing $\alpha$ and $\beta$ with their estimation $\hat{\alpha}$ and $\hat{\beta}$.

In conclusion, the complete process of the proposed active sampling method is given in Algorithm \ref{my algorithm}.

\begin{algorithm}
\caption{Active sampling algorithm for approximately bandlimited graph signals}
\label{my algorithm}
\begin{algorithmic}[1]
\STATE $t=0$, $M=0$
\STATE $\bm{\Psi}(0)=$ empty matrix, $\bm{y}_s(0)=$ empty vector
\STATE Initialize $\hat{\alpha}(0)$, $\hat{\beta}(0)$
\STATE Choose the first sampling node index $n$ arbitrarily
\STATE $\bm{\psi}^T(1)=\bm{e}_n^T$
\WHILE{$M<M_{max}$}
\STATE $t\gets t+1$
\STATE Sample the selected node and obtain observation $y(t)$
\vspace{-1em}
\STATE $\bm{\Psi}(t)=\left(\begin{matrix}\bm{\Psi}(t-1)\\\bm{\psi}^T(t)\end{matrix}\right)$, $\bm{y}_s(t)=\left(\begin{matrix}\bm{y}_s(t-1)\\y(t)\end{matrix}\right)$
\STATE $M\gets M+1$
\STATE $k=0$, $\hat{\alpha}_0=\hat{\alpha}(t-1)$, $\hat{\beta}_0=\hat{\beta}(t-1)$
\WHILE{$\hat{\alpha}$, $\hat{\beta}$ not converge}
  \STATE $k\gets k+1$
  \STATE Update $\bm{\mu}_{k-1}$, $\bm{C}_{k-1}$ by (\ref{E mean})\:(\ref{E covariance})
  \STATE Update $\hat{\alpha}_k$, $\hat{\beta}_k$ by (\ref{M alpha})\:(\ref{M beta})
\ENDWHILE
\STATE $\hat{\alpha}(t)=\hat{\alpha}_k$, $\hat{\beta}(t)=\hat{\beta}_k$
\STATE Update the posterior distribution by (\ref{posterior mean})\:(\ref{posterior covariance})
\STATE Design the next sampling node by (\ref{uncertainty sampling})
\IF{the stopping condition in (\ref{stopping condition}) is reached}
\STATE \textbf{break}
\ENDIF
\ENDWHILE
\STATE Ouput the MMSE estimation as (\ref{posterior mean})
\end{algorithmic}
\end{algorithm}

The sampling process continues until reaching the maximum sample size, or if the scalarized posterior covariance, normalized by the estimated signal energy, is less than a certain threshold $c>0$
\begin{equation}
\label{stopping condition}
\dfrac{\tr\left(\cov\left[\bm{f}\,|\,\bm{\Psi},\bm{y}_s,\hat{\alpha},\hat{\beta}\right]\right)}{{\hat{\bm{f}}}^T\hat{\bm{f}}}\leq c,
\end{equation}
which means the signal estimation is reliable enough.

\section{Simulation}
\label{sec4 simulation}
In this section, the proposed active sampling algorithm is applied to various synthetic approximately bandlimited graph signals to evaluate its performance.

\begin{figure}[t]
\centering
\subfloat[frequency response of $\bm{H}_1$]{
\includegraphics[width=0.20\textwidth]{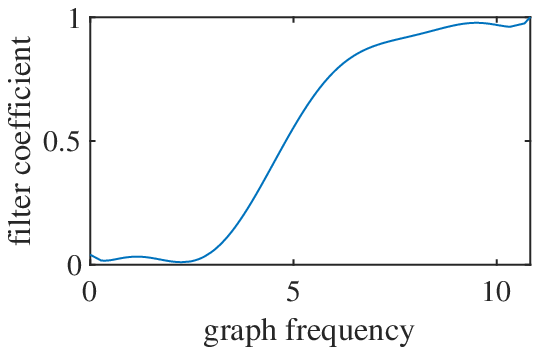}
\label{H1}}
\subfloat[frequency response of $\bm{H}_2$]{
\includegraphics[width=0.20\textwidth]{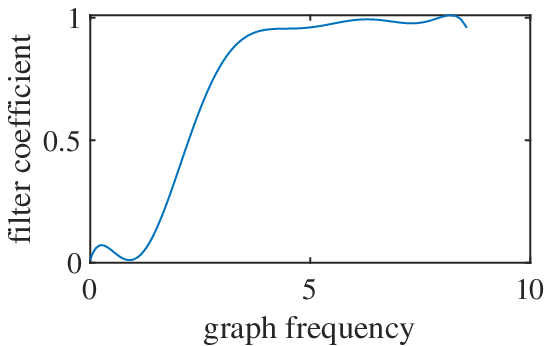}
\label{H2}}\\
\centering
\subfloat[GS1]{
\includegraphics[width=0.21\textwidth]{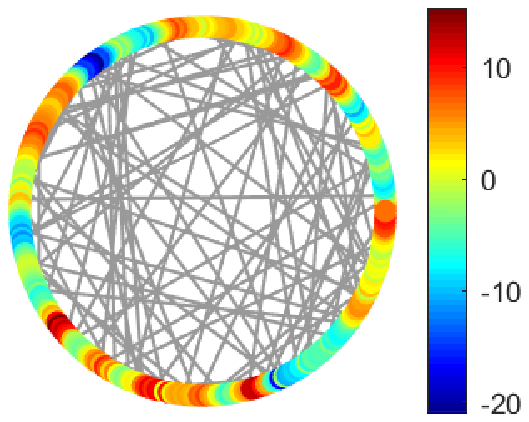}
\label{GS1}}
\subfloat[GS2]{
\includegraphics[width=0.21\textwidth]{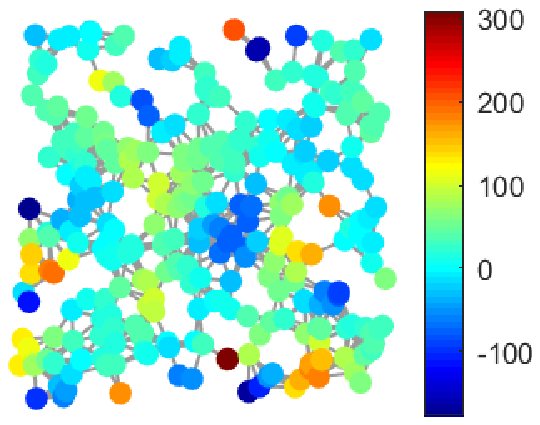}
\label{GS2}}\\
\centering
\subfloat[Spectrum of GS1]{
\includegraphics[width=0.20\textwidth]{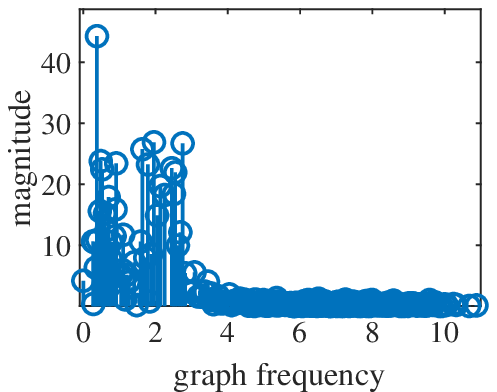}
\label{spectrum1}}
\subfloat[Spectrum of GS2]{
\includegraphics[width=0.20\textwidth]{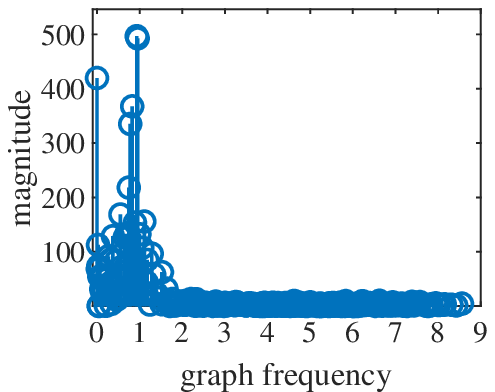}
\label{spectrum2}}
\caption{Graph signals for simulation and their spectrums}
\label{simulation GSs}
\end{figure}

Two representative graphs are used in our experiments: G1: a small world graph generated from the Watts-Strogatz model \cite{watts1998collective} with 300 nodes, mean node degree 6 and rewiring probability 0.1, and G2: a random geometric graph \cite{dall2002random,perraudin2014gspbox} with 300 vertices randomly placed in a 1 by 1 square and edge weights assigned via a Gaussian kernel
\begin{equation}
w_{ij}=\left\lbrace\begin{array}{ll}
\exp\left(\dfrac{d(i,j)^2}{\sigma^2}\right),&\text{if }d(i,j)\leq r\\
0,&\text{otherwise}
\end{array}\right.
\end{equation}
where $w_{ij}$ denotes the weight between node $i$ and node $j$, $d(i,j)$ denotes the Euclid distance between them, $r=0.1$ and $\sigma=0.05$.

Signals are generated from the prior distribution (\ref{prior}), where $\bm{H}$ is designed to be FIR \cite{shuman2011chebyshev}, the frequency response of which is displayed in Fig.\:(\ref{H1})\:(\ref{H2}), and $\alpha$ is set to be 10 and 0.1 for G1 and G2 respectively. Sampling noise is additive i.i.d Gaussian (\ref{noise 1}). For each graph, we pick two different $\beta$ such that the signal-noise ratios (SNR) are 15dB and 10dB. 100 signals are generated for each scenario to evaluate the average performance of each method. Fig.\:\ref{simulation GSs} shows two instances of the graph signals used in our simulations.

Note that existing design-of-experiments(DOE)-type sampling methods cannot apply to the scenario where $\alpha$ and $\beta$ are unknown. The proposed algorithm is mainly compared with random sampling (and estimating signal $\bm{f}$ in the same way as proposed) in the simulations. In fact, there are also non-statistical approaches such as Perraudin's non-uniform sampling based on local uncertainty \cite{perraudin2018global} (M1) and Anis's heuristic algorithm to maximize the cut-off frequency \cite{anis2016efficient} (M2), the results of which are also displayed here.

\begin{figure}[t]
\centering
\subfloat[G1, $\alpha=10$, $\text{SNR}=15$dB]{
\includegraphics[width=0.22\textwidth]{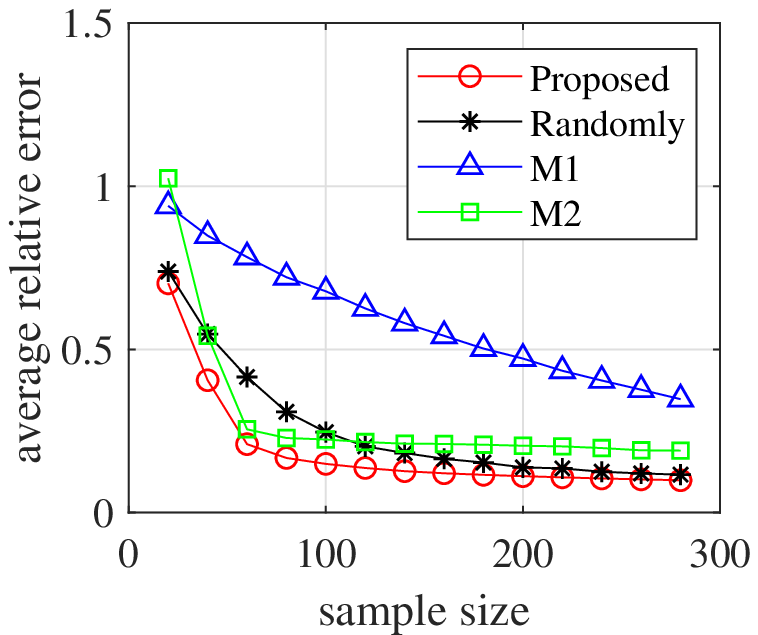}
\label{result1}}
\subfloat[G1, $\alpha=10$, $\text{SNR}=10$dB]{
\includegraphics[width=0.22\textwidth]{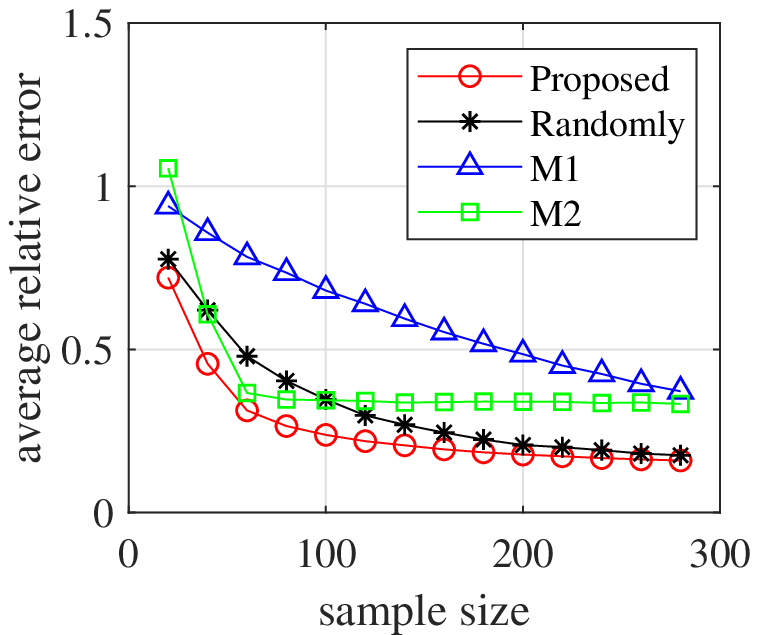}
\label{result2}}\\
\centering
\subfloat[G2, $\alpha=0.1$, $\text{SNR}=15$dB]{
\includegraphics[width=0.22\textwidth]{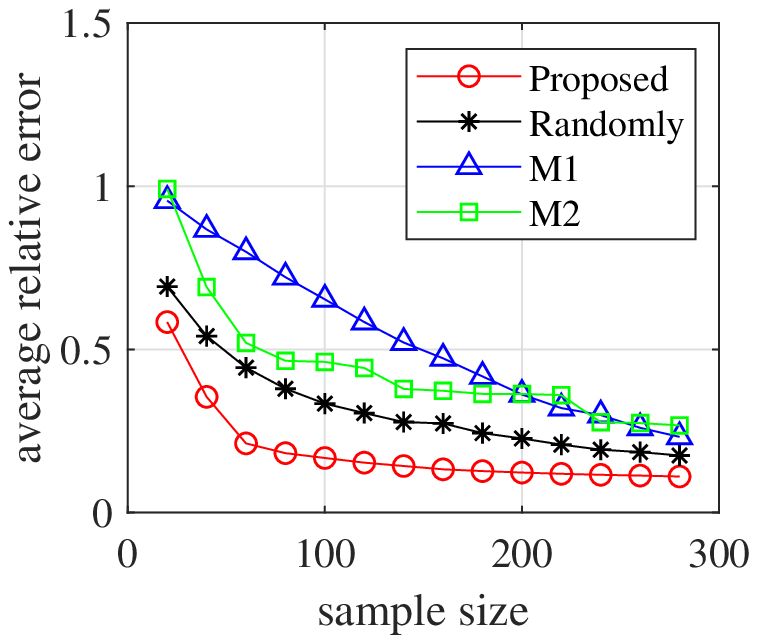}
\label{result3}}
\subfloat[G2, $\alpha=0.1$, $\text{SNR}=10$dB]{
\includegraphics[width=0.22\textwidth]{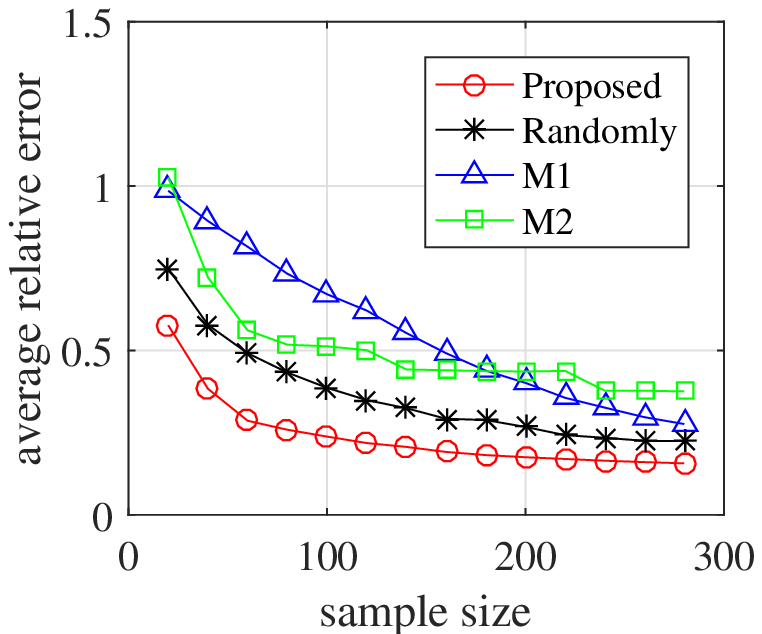}
\label{result4}}
\caption{Simulation results}
\label{simulation results}
\end{figure}

Fig.\:\ref{simulation results} displays the simulation results, where relative estimation error is defined as
\begin{equation}
err=\dfrac{\Vert\hat{\bm{f}}-\bm{f}\Vert_2}{\Vert\bm{f}\Vert_2}.
\end{equation}
We can see that the performance of the proposed algorithm is significantly better than random sampling regardless of the graph type, signal smoothness and noise power within a certain range. The proposed method requires less samples to reach a given estimation accuracy. This validates the rationality and effectiveness of our effort to involve active learning to gradually estimate the parameters in the sampling process and decide the subsequent sampling nodes based on the latest model.

\section{Conclusion}
\label{sec5 conclusion}
In this paper, an active sampling algorithm is proposed for approximately bandlimited graph signals without prior knowledge about the exact distribution of the signal and noise. By implementing this active sampling strategy, possibly less nodes are required to be sampled to reach the same estimation accuracy.

A more general case may be considered in our further research where the high-pass graph filter $\bm{H}$ in the signal prior is also unknown. The active sampling of bandlimited graph signals with unknown bandwidth will be studied as well.

\section{Acknowledgment}
This work was supported by the NSF of China (No. 61501124) and the National Key R\&D Program of China (No. 213).

\bibliographystyle{IEEEbib}
\bibliography{References}

\begin{thebibliography}{10}

\bibitem{shuman2013emerging}
David~I Shuman, Sunil~K Narang, Pascal Frossard, Antonio Ortega, and Pierre
  Vandergheynst,
\newblock ``The emerging field of signal processing on graphs: Extending
  high-dimensional data analysis to networks and other irregular domains,''
\newblock {\em IEEE Signal Process. Mag.}, vol. 30, no. 3, pp. 83--98, 2013.

\bibitem{ortega2018graph}
Antonio Ortega, Pascal Frossard, Jelena Kova{\v{c}}evi{\'c}, Jos{\'e}~MF Moura,
  and Pierre Vandergheynst,
\newblock ``Graph signal processing: Overview, challenges, and applications,''
\newblock {\em Proc. IEEE}, vol. 106, no. 5, pp. 808--828, 2018.

\bibitem{pesenson2008sampling}
Isaac Pesenson,
\newblock ``Sampling in paley-wiener spaces on combinatorial graphs,''
\newblock {\em Transactions of the American Mathematical Society}, vol. 360,
  no. 10, pp. 5603--5627, 2008.

\bibitem{narang2013signal}
Sunil~K Narang, Akshay Gadde, and Antonio Ortega,
\newblock ``Signal processing techniques for interpolation in graph structured
  data,''
\newblock in {\em Proc. IEEE Int. Conf. Acoust., Speech, Signal Process.
  (ICASSP)}, 2013, pp. 5445--5449.

\bibitem{anis2014towards}
Aamir Anis, Akshay Gadde, and Antonio Ortega,
\newblock ``Towards a sampling theorem for signals on arbitrary graphs,''
\newblock in {\em Proc. IEEE Int. Conf. Acoust., Speech, Signal Process.
  (ICASSP)}, 2014, pp. 3864--3868.

\bibitem{chen2015discrete}
Siheng Chen, Rohan Varma, Aliaksei Sandryhaila, and Jelena Kova{\v{c}}evi{\'c},
\newblock ``Discrete signal processing on graphs: Sampling theory,''
\newblock {\em IEEE Trans. Signal Process.}, vol. 63, no. 24, pp. 6510--6523,
  2015.

\bibitem{marques2016sampling}
Antonio~G Marques, Santiago Segarra, Geert Leus, and Alejandro Ribeiro,
\newblock ``Sampling of graph signals with successive local aggregations.,''
\newblock {\em IEEE Trans. Signal Process.}, vol. 64, no. 7, pp. 1832--1843,
  2016.

\bibitem{xie2017design}
Xuan Xie, Hui Feng, Junlian Jia, and Bo~Hu,
\newblock ``Design of sampling set for bandlimited graph signal estimation,''
\newblock in {\em Proc. IEEE Global Conf. Signal Inf. Process. (GlobalSIP)},
  2017, pp. 653--657.

\bibitem{anis2016efficient}
Aamir Anis, Akshay Gadde, and Antonio Ortega,
\newblock ``Efficient sampling set selection for bandlimited graph signals
  using graph spectral proxies,''
\newblock {\em IEEE Trans. Signal Process.}, vol. 64, no. 14, pp. 3775--3789,
  2016.

\bibitem{chen2016signal}
Siheng Chen, Rohan Varma, Aarti Singh, and Jelena Kova{\v{c}}evi{\'c},
\newblock ``Signal recovery on graphs: Fundamental limits of sampling
  strategies,''
\newblock {\em IEEE Trans. Signal Inf. Process. Netw.}, vol. 2, no. 4, pp.
  539--554, 2016.

\bibitem{perraudin2018global}
Nathanael Perraudin, Benjamin Ricaud, David~I Shuman, and Pierre Vandergheynst,
\newblock ``Global and local uncertainty principles for signals on graphs,''
\newblock {\em APSIPA Transactions on Signal and Information Processing}, vol.
  7, 2018.

\bibitem{settles2012active}
Burr Settles,
\newblock ``Active learning,''
\newblock {\em Synthesis Lectures on Artificial Intelligence and Machine
  Learning}, vol. 6, no. 1, pp. 1--114, 2012.

\bibitem{zhu2003combining}
Xiaojin Zhu, John Lafferty, and Zoubin Ghahramani,
\newblock ``Combining active learning and semi-supervised learning using
  gaussian fields and harmonic functions,''
\newblock in {\em Proc. ICML 2003 workshop on the continuum from labeled to
  unlabeled data in machine learning and data mining}, 2003, vol.~3.

\bibitem{long2008graph}
Jun Long, Jianping Yin, Wentao Zhao, and En~Zhu,
\newblock ``Graph-based active learning based on label propagation,''
\newblock in {\em Proc. International Conference on Modeling Decisions for
  Artificial Intelligence}, 2008, pp. 179--190.

\bibitem{zhou2014active}
Jin Zhou and Shiliang Sun,
\newblock ``Active learning of gaussian processes with manifold-preserving
  graph reduction,''
\newblock {\em Neural Computing and Applications}, vol. 25, no. 7-8, pp.
  1615--1625, 2014.

\bibitem{berberidis2017active}
Dimitris Berberidis and Georgios~B Giannakis,
\newblock ``Active sampling for graph-aware classification,''
\newblock in {\em Proc. IEEE Global Conf. Signal Inf. Process. (GlobalSIP)},
  2017, pp. 648--652.

\bibitem{barber2012bayesian}
David Barber,
\newblock {\em Bayesian reasoning and machine learning},
\newblock Cambridge University Press, 2012.

\bibitem{gadde2015probabilistic}
Akshay Gadde and Antonio Ortega,
\newblock ``A probabilistic interpretation of sampling theory of graph
  signals,''
\newblock in {\em Proc. IEEE Int. Conf. Acoust., Speech, Signal Process.
  (ICASSP)}. IEEE, 2015, pp. 3257--3261.

\bibitem{shuman2011chebyshev}
David~I Shuman, Pierre Vandergheynst, and Pascal Frossard,
\newblock ``Chebyshev polynomial approximation for distributed signal
  processing,''
\newblock in {\em International Conference on Distributed Computing in Sensor
  Systems and Workshops (DCOSS)}, 2011, pp. 1--8.

\bibitem{shi2015infinite}
Xuesong Shi, Hui Feng, Muyuan Zhai, Tao Yang, and Bo~Hu,
\newblock ``Infinite impulse response graph filters in wireless sensor
  networks,''
\newblock {\em IEEE Signal Process. Lett.}, vol. 22, no. 8, pp. 1113--1117,
  2015.

\bibitem{kay1993fundamentals}
Steven~M Kay,
\newblock {\em Fundamentals of statistical signal processing, volume I:
  Estimation theory},
\newblock Prentice Hall, 1993.

\bibitem{watts1998collective}
Duncan~J Watts and Steven~H Strogatz,
\newblock ``Collective dynamics of ‘small-world’networks,''
\newblock {\em Nature}, vol. 393, no. 6684, pp. 440, 1998.

\bibitem{dall2002random}
Jesper Dall and Michael Christensen,
\newblock ``Random geometric graphs,''
\newblock {\em Phys. Rev. E}, vol. 66, no. 1, pp. 016121, 2002.

\bibitem{perraudin2014gspbox}
Nathana{\"e}l Perraudin, Johan Paratte, David Shuman, Lionel Martin, Vassilis
  Kalofolias, Pierre Vandergheynst, and David~K Hammond,
\newblock ``{GSPBOX}: A toolbox for signal processing on graphs,''
\newblock {\em arXiv preprint arXiv: 1408.5781}, 2014.

\end{thebibliography}

\end{document}